\documentclass[reprint,aps]{revtex4-2}
\usepackage{graphicx}% Include figure files
\usepackage{dcolumn}% Align table columns on decimal point
\usepackage{bm}% bold math
\usepackage{amsmath,amssymb,amsthm,mathrsfs,amsfonts,dsfont}
\usepackage{array,makecell,float,mathpazo,exscale,varwidth,paralist,bbm,mathtools}
\usepackage{bbold}
\usepackage{tikz-cd}

\usepackage{braket}

\begin{document}

\preprint{APS/123-QED}

\title{Anomalous non-classicality via correlative and entropic Bell inequalities}

\author{Sabiha Durucan}
\affiliation{Fakult\"at f\"ur Physik, Universität Bielefeld, 33615 Bielefeld, Germany}

\author{Alexei Grinbaum}%
\affiliation{%
CEA-Saclay, IRFU/Larsim, 91191 Gif-sur-Yvette, France
}%

\date{\today}% It is always \today, today,

\begin{abstract}

Entropic Bell inequalities witness contextual probability distributions on sets of jointly measurable observables. We find that their violation does not entail a violation of the correlative Bell inequality for certain parameter values. This anomaly between the entropic and correlative measures of contextuality helps to precisely determine the type of non-classical resource. We determine its numerical bounds inviting their experimental verification. Among the permutations of observables that keep this anomaly in place, we identify the `exotic' type that gives rise to non-classical resources under all non-equivalent operational party assignments in a device-independent approach.
\end{abstract}

\maketitle

\section{Introduction}

It is known that maximally entangled states do not achieve the maximum violation of certain Bell inequalities~\cite{RevNonLoc}. This ``anomaly of non-locality''~\cite{ScaraniAnomaly} indicates that entanglement and non-locality are different non-classical resources.
Quantum contextuality is yet another non-classical resource~\cite{KochSpeck,PhysRevLett.119.120505,BartlettResource}. When it is witnessed by the violation of entropic inequalities~\cite{BraunCaves}, the measure of contextuality diverges from the assessment of non-classicality based on the correlative Bell inequalities. By exploring this divergence numerically in the bipartite scenario, we provide the first thorough comparison of the entropic and correlative measures of contextuality.
% The latter differ from the assessment of non-classicality via correlative inequalities, exhibiting yet another anomaly. Here, we explore this %anomaly numerically and compute }

Fritz and Chaves found that the presence of contextuality measured entropically (hereinafter e-contextuality) should necessarily imply the presence of contextuality measured via the correlations (hereinafter c-contextuality)~\cite{FritzChaves}. Yet measurements leading to maximum violation of the entropic inequality are not the ones that produce the maximum violation of the correlative Clauser-Horne-Schimony-Holt ($CHSH$) inequality.
Even more surprisingly, there exist \textit{anomalous} probability distributions that violate the entropic inequality but respect the correlative one: for certain ranges of measurement parameters,\  e-contextuality can be detected \textit{without} c-contextuality. \\

The anomaly can be explained by the failure of entropic measures to distinguish between perfectly correlated and anti-correlated variables. We also see its further conceptual significance. Hailed as a ``very important recent development''~\cite{popescu2014}, the device-in\-depen\-dent approach consists in describing an experiment by specifying its input and output as strings in alphabets of finite cardinality~\cite{Lang,baumeler_2}.
A party, conventionally conceived of by delimiting its spatial location or by identifying the physical system on which it acts, is now defined by associating algebraic input and output variables over many runs of the experiment. Different assignments lead to non-equivalent settings, some of which may contain causally non-independent or non-local parties; for each assignment, however, it is possible to formulate Bell inequalities that measure non-classicality of the corresponding probability distribution. We demonstrate possible variations in violating the entropic and correlative inequalities as one reassigns inputs and outputs to the parties, leading to the disappearance of the anomaly in some cases. The anomaly can also persist: we give examples of probability distributions that remain contextual under any party assignment. These situations suggest that in the device-independent approach there exist resources that are non-classical in all non-equivalent operational settings.

After introducing the framework in section \ref{sect:math}, we analyze the difference between correlative and entropic measures of non-classicality
in section \ref{section:results}. In section \ref{sect:perm}, we explore `permissible', `forbidden', and `exotic' permutations in a device-independent approach.

\section{Mathematical framework}\label{sect:math}
Suppose that two parties, $A$ and $B$, share a quantum state $\ket{\psi}$. At each run of the experiment $A$ and $B$ choose a binary input, resp. $x,y\in \{0,1\}$, and measure a corresponding observable, resp. $A_x,B_y$, obtaining a binary outcome, resp. $a,b\in \{0,1\}$. The presence of non-classical correlations between $A$ and $B$ is witnessed by a violation of the Bell inequality~\cite{Bell1} in the $CHSH$ form~\cite{CHHS}
\begin{equation}
CHSH=\braket{A_0B_0}+\braket{A_0B_1}+\braket{A_1B_0}-\braket{A_1B_1} \leq 2.
\label{eq:chsh-original}\end{equation}

Another measure of non-classicality is based on entropic inequalities. In the bipartite Bell scenario, two observables from different parties, $A_x$ and $B_y$, are jointly measurable for all values of $x$ and $y$, while two observables of the same party, e.g. $A_0$ and $A_1$, are not. Define a marginal scenario consisting of all sets of jointly measurable observables~\cite{FritzChaves1}. This scenario is called non-contextual if the joint probability $P(a,b|x,y)$ can be written as a function of hidden variable $\lambda$ with probability distribution $\rho(\lambda)$:
\begin{equation}
    P(a,b\ |\ x,y)=\sum_\lambda \rho(\lambda)P(a|x,\lambda)P(b|y,\lambda).\label{eq:hidden}
\end{equation}
%Any scenario with a probability distribution that cannot be so expressed is called contextual.
Based on marginal scenarios, Fritz and Chaves proved that $P(a,b|x,y)$ is e-contextual if and only if the following entropic inequality is violated~\cite{FritzChaves}
\begin{align}
CHSH_E = H(A_1,B_1)+H(A_0)+H(B_0)- \nonumber\\ H(A_0,B_0)-H(A_0,B_1)-H(A_1,B_0)\leq 0,\label{eq:chsh_E}
\end{align}
where $H$ denotes respective Shannon entropies.

Choose an entangled pure state $\ket{\psi}\ =\cos\alpha \ket{00}+ \sin \alpha \ket{11}$. The maximum violation of inequalities (\ref{eq:chsh-original}) and (\ref{eq:chsh_E}) can be obtained at $\alpha=\pi/4$ by parties $A$ and $B$ performing measurements in the $Y$-$Z$ plane at angles $\theta_x$ and $\theta_y ^\prime$ respectively:
\begin{align*}
A_x\ =\ \sin\theta_x \sigma_Y + \cos \theta_x \sigma_Z \\
B_y\ =\ \sin\theta_{y}^\prime \sigma_Y + \cos \theta_{y}^\prime\sigma_Z,
\end{align*}
where $\sigma_Y$ and $\sigma_Z$ are Pauli matrices. Strong symmetries at $\alpha=\pi/4$ imply $P(1,1|x,y)=P(0,0|x,y)$, $P(0,1|x,y)=P(1,0|x,y)$, and $H(A_x)=H(B_y)=1$. Quantum theory allows for any values $\theta_x,\theta_y'\in[0,2\pi]$.
The maximum violation of inequality (\ref{eq:chsh-original}) over this entire range is equal to the Tsirelson bound $2\sqrt{2}$~\cite{Tsirel}, while the violation of (\ref{eq:chsh_E}) reaches at most $CHSH_E\simeq 0.237$ ~\cite{FritzChaves}. Note that the measurement angles $\theta_x,\theta^\prime _y$ that maximise $CHSH$ yield non-maximal values of $CHSH_E$, and vice versa (Table~\ref{tab:chshe_max_vio}).

\begin{table}%[h]
    \centering
    \begin{tabular}{|c|c|c|c|c|c|}\hline
       $\theta_0$    &$\theta_1$   &$\theta_0'$  &$\theta_1'$  &$CHSH$ &$CHSH_E$  \\
       \hline
        2.070&1.466&1.372&0.769&2.248&0.2369\\
        2.709&2.106&0.739&0.125&2.250&0.2368\\ \hline
        1.316&2.894&1.033&2.606&2.828&-1.205\\
        2.050&0.486&1.877&0.294&2.828&-1.210\\ \hline
    \end{tabular}
    \caption{Examples of measurement settings leading to the maximum violation of the $CHSH$ and $CHSH_E$ inequalities.}
    \label{tab:chshe_max_vio}
\end{table}

\section{Discrepancy in evaluating non-classicality}\label{section:results}
\begin{figure}[h!]
    \centering
    \includegraphics[width=0.48\textwidth]{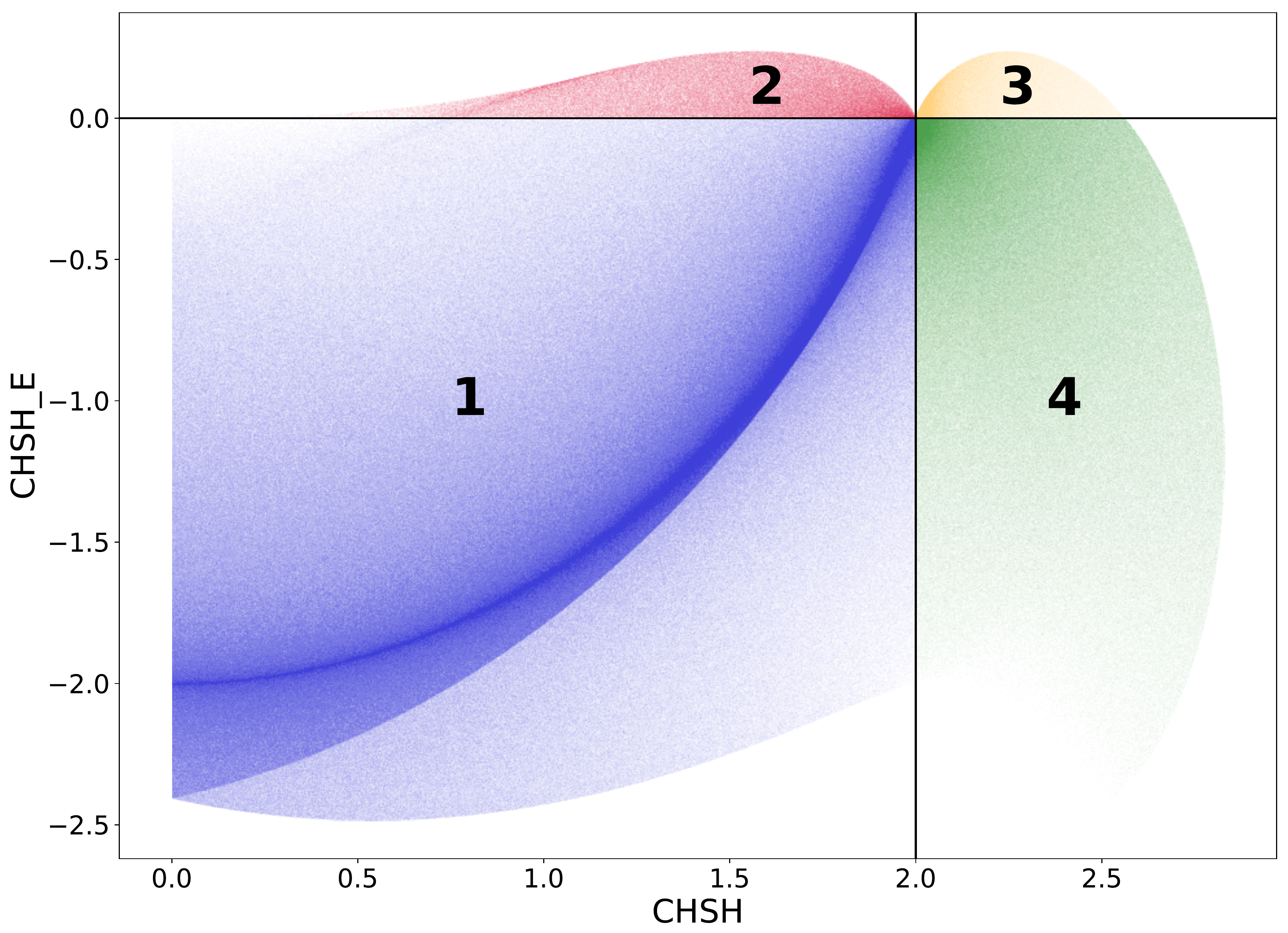}
    \caption{Values of $CHSH$ and $CHSH_E$ for 10 million measurement settings. Non-classicality is detected by e-contextuality if $CHSH_E>0$ and/
    or by c-contextuality %non-classical correlations
    if $CHSH>2$.}
    \label{fig:chshvschshe}
\end{figure}

Correlative and entropic $CHSH$ inequalities (\ref{eq:chsh-original}) and (\ref{eq:chsh_E}) capture non-classicality in distinct ways. Figure~\ref{fig:chshvschshe} demonstrates this discrepancy on 10 million measurement settings given by a set of four angles $\theta_x,\theta_y^\prime$.
Zone 1 contains all settings that do not violate either inequality, i.e. probability distributions that are non-contextual. Zone 3 contains non-classical settings that violate both inequalities, while in zones 2 and 4 only one of the inequalities, resp. $CHSH_E$ and $CHSH$, detects non-classicality. For $\alpha=\pi/4$, approximately 82.5\% of measurements belong to zone 1 and the remaining 17.5\% demonstrate at least some non-classical behavior (Table~\ref{tab:zonedensities}).
\begin{table}%[h!]
    \centering
    \begin{tabular}{|c|c|c|c|}\hline
         Zone 1&Zone 2&  Zone 3&  Zone 4 \\ \hline
          82.5$\%$& 1.4$\%$& 2.9$\%$& 13.2\\ \hline
    \end{tabular}
    \caption{Proportions of 10 million measurement settings in zones 1-4.}
    \label{tab:zonedensities}
\end{table}

\begin{figure}[h]
    \centering
    \includegraphics[width=0.48\textwidth]{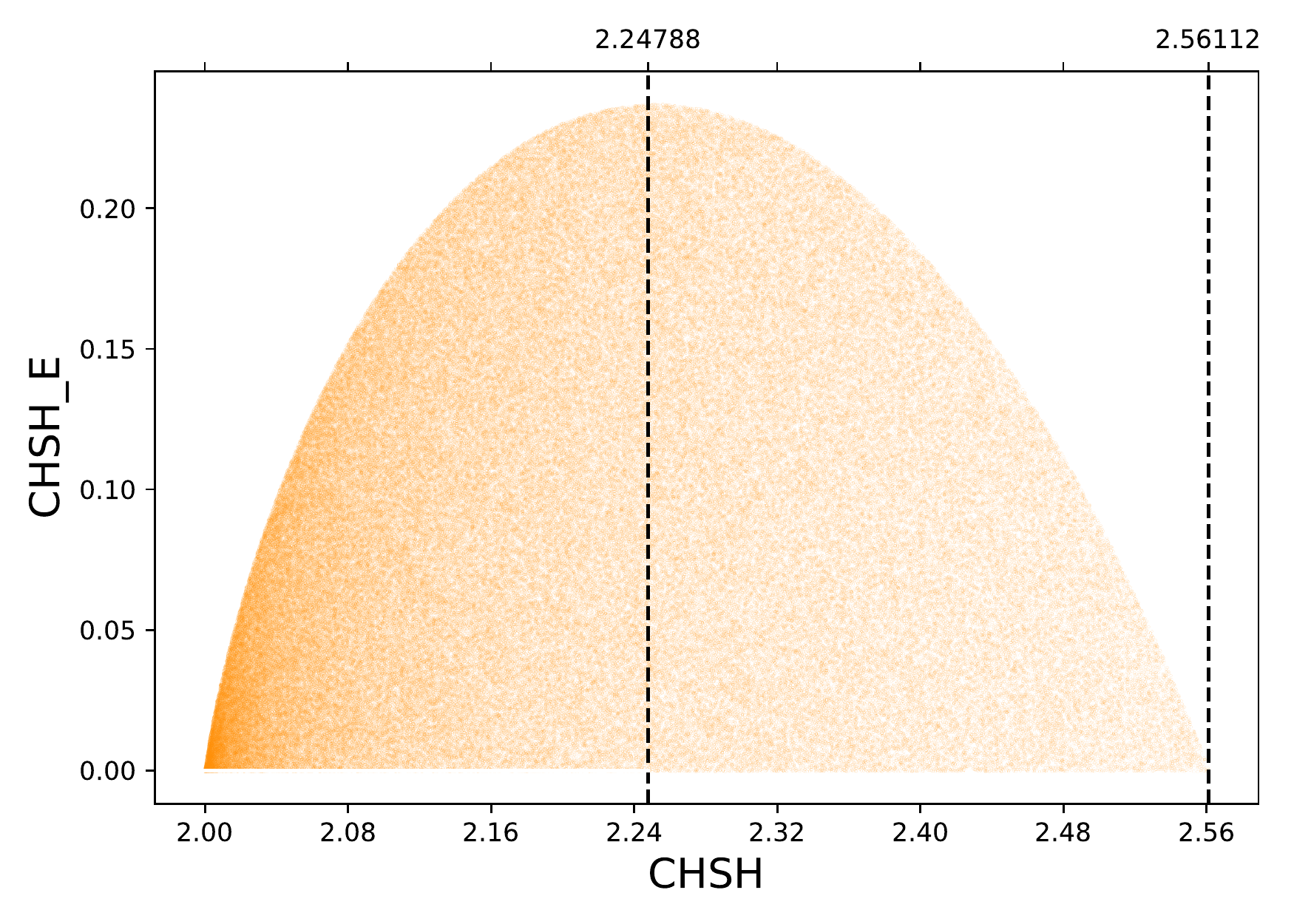}
    \caption{Zone 3 of the graph in Figure~\ref{fig:chshvschshe}. Inequalities (\ref{eq:chsh-original}) and (\ref{eq:chsh_E}) are both violated.}
    \label{fig:parabola1}
\end{figure}

Zone 3 is shown in detail on Figure~\ref{fig:parabola1}. It vanishes at $CHSH\simeq 2.561$, corresponding to the maximum violation of the correlative inequality~(\ref{eq:chsh-original}) that can be reached in an e-contextual setting. The $CHSH$ values beyond this bound, in particular the ones close to the Tsirelson bound, can only be witnessed by e-noncontextual but c-contextual probability distributions. Reproducing this analysis for $0<\alpha < \pi/4$, we find that zone 3 decreases in size more rapidly than zones 2 and 4, implying that the violations of non-classicality in the correlative and entropic forms are more often witnessed separately.

Zone 2 illustrates an effect similar to the ``anomaly of non-locality''~\cite{ScaraniAnomaly}. If one considers the maximum values of $CHSH$ and $CHSH_E$ over all theoretically allowed measurement angles, $\theta_x,\theta_y^\prime\in [0,2\pi]$, then violating inequality~(\ref{eq:chsh-original}) is a necessary (but not a sufficient) condition for violating inequality (\ref{eq:chsh_E}).
In zone 2, however, only inequality~(\ref{eq:chsh_E}) is violated. Measurement settings that belong to this zone yield anomalous e-contextuality \textit{without} c-contextuality in the bipartite setting.

\begin{figure}[!htb]
    \centering
    \includegraphics[width=0.48\textwidth]{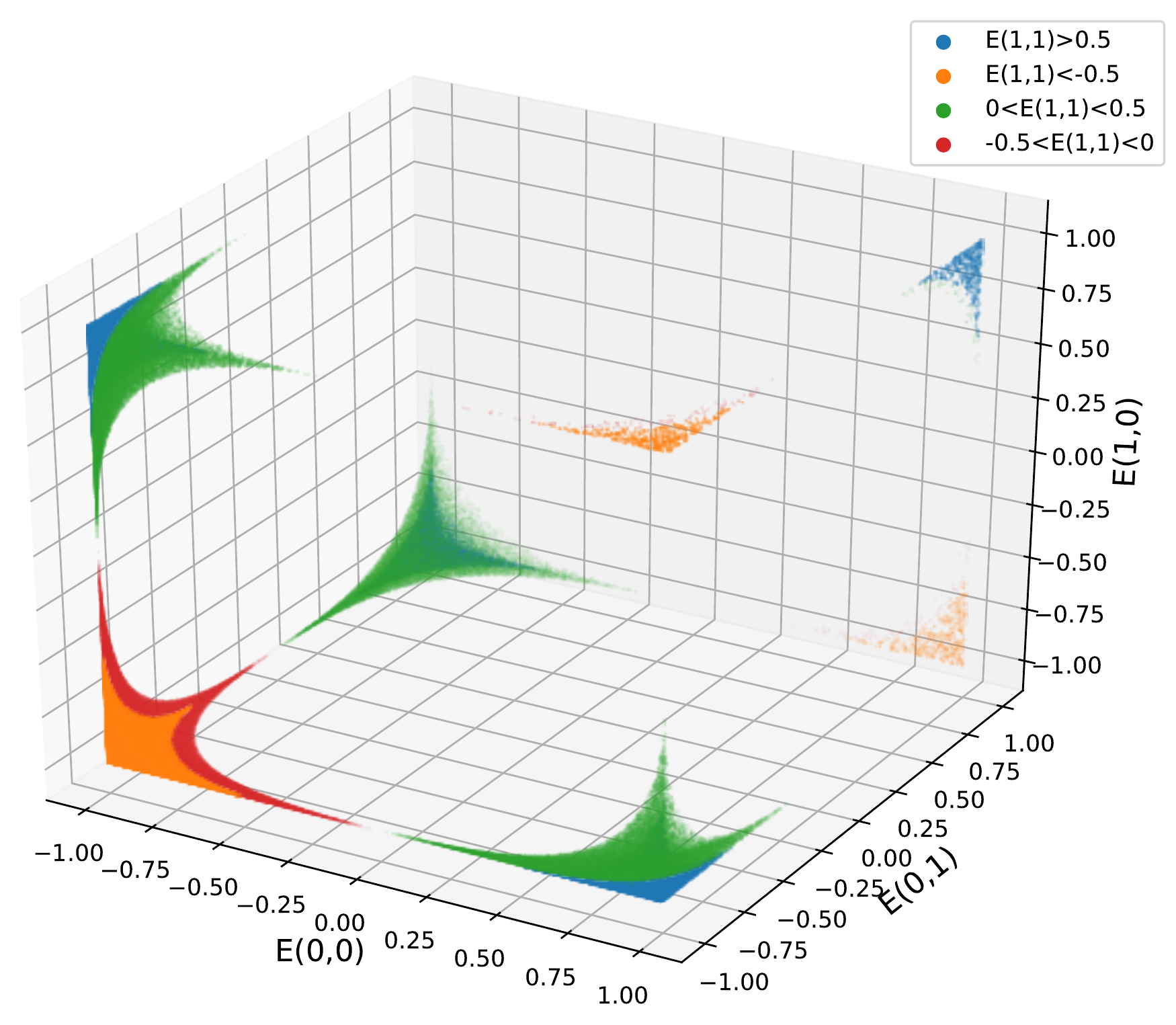}
    \includegraphics[width=0.48\textwidth]{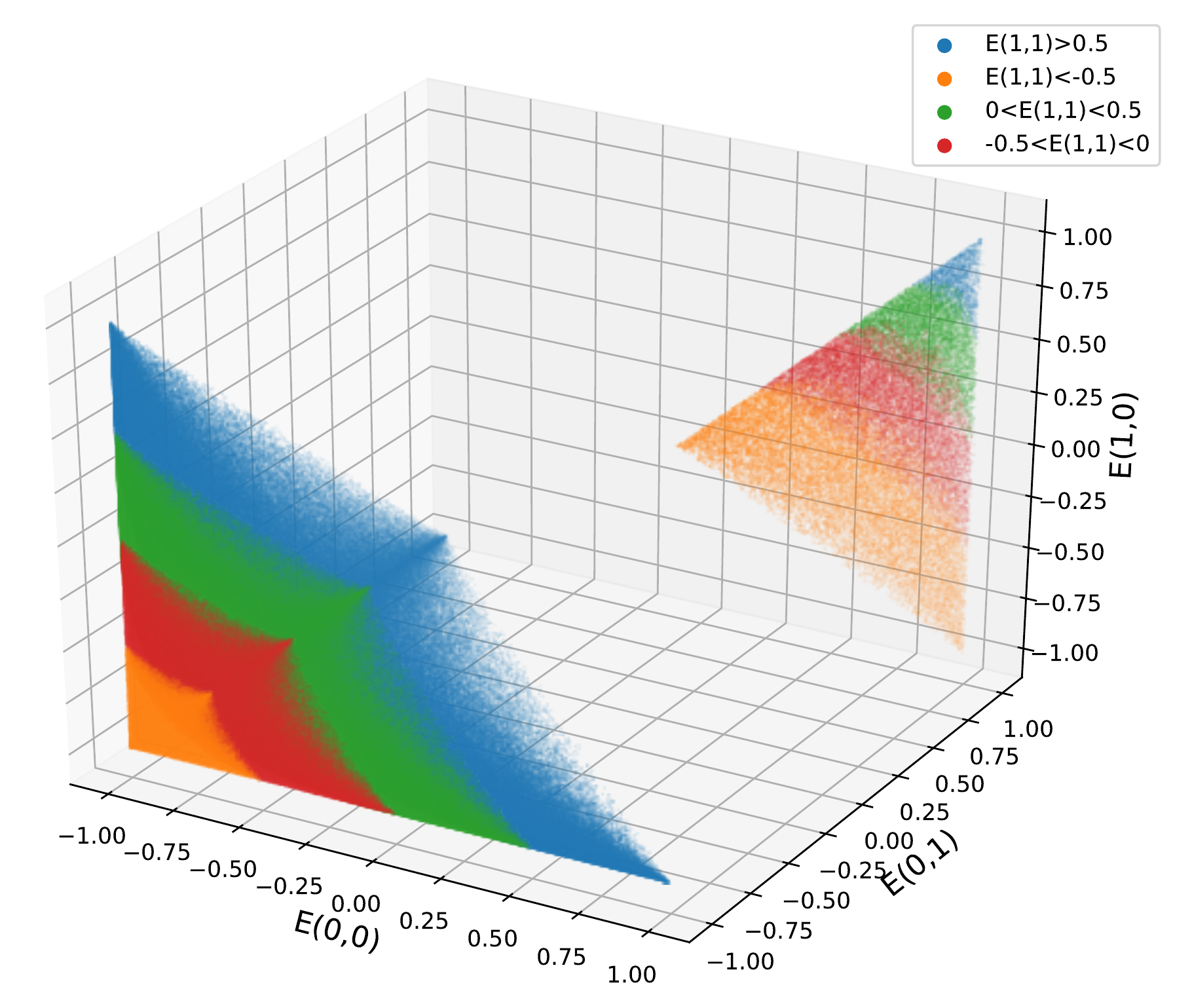}
    \caption{Top: Correlators of the probability distributions that violate inequality~(\ref{eq:chsh_E}). Violations are maximal in the `corners' of high correlation or anti-correlation. Bottom: Correlators of the probability distributions that violate inequality~(\ref{eq:chsh-original}). %The Tsirelson bound is reached near the point
    }
    \label{fig:corrchshe}
\end{figure}

This anomalous behavior is due to the failure of entropic measures to distinguish between perfectly correlated and anti-correlated variables. Its origin can be illustrated by plotting the correlators $E(x,y)=P(a=b|x,y)-P(a\neq b|x,y)$ for $CHSH_E>0$ (Figure~\ref{fig:corrchshe}). One would expect an entropic inequality to be violated when different variables are maximally uncorrelated, i.e. relative entropies are high. In reality, maximum violations of the entropic inequality (\ref{eq:chsh_E}) occur close to all but one `corners' of highly correlated and/or highly anti-correlated outputs. Note that the correlative inequality~(\ref{eq:chsh-original}) remains sensitive to this distinction and cannot be violated in some of these `corners'.

\section{Permissible, forbidden, and exotic permutations}\label{sect:perm}

Fritz and Chaves discuss permutations that transform an entropic contextuality inequality into an equivalent one~\cite{FritzChaves}. Similarly, one can consider permutations that lead to equivalent correlative inequalities. These `permissible' permutations include permutations of the parties, permutations of the observables of some party, or their arbitrary combinations. For example, exchanging $B_0$ with $B_1$ transforms eq. (\ref{eq:chsh_E}) into an equivalent inequality:
%An equivalent inequality is obtained by permuting signs in the original formula, for example exchanging $B_0$ with $B_1$ transforms (\ref{eq:chsh_E}) into
\begin{align}
CHSH_{E}^\prime=H(A_0)+H(B_1)-H(A_0B_0)-\nonumber \\ H(A_0B_1)+H(A_1B_0)-H(A_1B_1)\leq 0.\label{eq:chsh_E2}
\end{align}
In any given marginal scenario only one among four such sign-permuted inequalities can witness non-classicality. %too early/confusing to mention this already?

In a device-independent approach~\cite{grin_devindep}, suppose that each run of the experiment is characterized by a binary string of length four. A homomorphism on these strings, mapping bit value of run $i$ on bit value of run $j$ over all runs, defines input and output variables, e.g. $x,y$ and $a,b$ respectively in the notation of Section~\ref{sect:math}. The identification of an input variable with an output variable, e.g. $x$ with $a$ and $b$ with $y$, provides a definition of parties $A$ and $B$. Different assignments lead to different interpretations of the experiment in terms of parties (`local observers that perform local operations'), while physically meaningful interpretations also require additional causal assumptions~\cite{ChavesCausal}, e.g. the no-signalling condition $P(a|x,y)=P(a|x),\;P(b|x,y)=P(b|y)$ typically used in the device-independent models~\cite{popescu}.

Since permissible permutations respect joint measurability, they result in eight equivalent measurement settings for a given definition of parties. Now, instead of observer-specific notation $A_x$ and $B_y$, consider four observables in a more general setting with no \textit{a priori} attribution to local observers. For each given set $\{X_i\}_{i=1\ldots 4}$, there exist three definitions of local parties that lead to different marginal scenarios:
\begin{description}
    \item[Class 1] Observer $A$ has $X_1,X_2$; observer $B$ has $X_3,X_4$;
    \item[Class 2] Observer $A$ has $X_1,X_4$; observer $B$ has $X_2,X_3$;
    \item[Class 3] Observer $A$ has $X_1,X_3$; observer $B$ has $X_2,X_4$.
\end{description}
These classes yield distinct party assignments, e.g. class 2 corresponds to $A$ having what was formerly known as ``$A_0$'' and ``$B_1$'', while $B$ has what had been labelled ``$A_1$'' and ``$B_0$'' in Section~\ref{sect:math}. The attribution of labels $A$ and $B$ to parties is arbitrary, for it is subject to a permissible permutation. The classes are conserved under permissible as well as `forbidden' permutations that do not preserve the entropic contextuality inequality but respect joint measurability, i.e. they remain within the same marginal scenario. An example of a `forbidden' permutation is given by exchanging $B_0$ and $B_1$ without switching from inequality~(\ref{eq:chsh_E}) to (\ref{eq:chsh_E2}).

Different classes are connected by `exotic' permutations that change the marginal scenario, for instance exchanging $A_0$ and $B_0$. A physical interpretation under exotic permutations may be complicated by the fact that some parties admit signaling or lack operational meaning as local observers, for example by exhibiting non-local properties in time~\cite{Oreshkov2018}. It is a legitimate concern to require that an assignment of parties be physically well-defined in order to be empirically realizable; here, we ask whether theoretically allowed probability distributions, connected by forbidden or exotic permutations, exhibit non-classical behavior in the correlative or entropic sense.

Under some, but not all, forbidden permutations the probability distributions change zones in Figure~\ref{fig:chshvschshe}. Three possible situations include staying in zone 1, switching between zone 1 and any other zone, or switching between zones 2 and 4. Thus a forbidden permutation, i.e. an inequivalent assignment of parties that nevertheless respects joint measurability of the observables, may erase non-classicality by bringing the probability distribution from any non-classical zone into zone 1. We find that there always exists precisely one such permutation that will change an anomalous e-contextual distribution without c-contextuality (zone 2) into a c-contextual distribution not detected by the entropic inequality (zone 4). This follows from the fact that no hidden-variable model~(\ref{eq:hidden}) exists for the setting that violates inequality (\ref{eq:chsh_E}), hence at least one forbidden permutation on the probability distribution that implements the same party assignment for inequality (\ref{eq:chsh-original}) will lead to its violation. Since only one among the non-equivalent sign-permuted inequalities is violated in a given marginal scenario, the forbidden permutation between zones 2 and 4 is unique.

Exotic permutations, contrary to the forbidden ones, can accommodate any change of zone in Figure~\ref{fig:chshvschshe}. Anomalous non-classicality can therefore be interpreted as an artifact of unfortunate choice of the setting, in particular of a problematic party assignment: although such an assignment might be physical in the sense of observer locality, it is problematic in the sense of containing a non-classical resource that can be dissolved by permutation.

Interestingly, there exist cases when all three non-equivalent scenarios, i.e. all non-equivalent party assignments on a set of inputs and outputs, are contextual in the correlative and entropic sense. In these cases even `exotic' permutations between classes $1-3$ do not alter non-classicality of the probability distribution. Two examples of such triple violation of inequality~(\ref{eq:chsh_E}) in zones 2 and 3 are shown in Table~\ref{tab:allfour}.
\vskip 1em

\begin{table}[h!]
    \centering
    \begin{tabular}{|c|c|c|c|c|c|c|}\hline
Zone&Angles&Class 1&Class 2&Class 3\\ \hline
    2&$\theta_0=0.40$,\;$\theta_1=3.02,$&$1.71$&$1.71$&$1.42$\\
    &$\theta_0^\prime=2.72$,\;$\theta_1^\prime=2.38$&$0.11$&$0.07$&$0.18$\\
\hline
    3&$\theta_0=1.97$,\;$\theta_1=1.31,$&$2.22$&$2.07$&$2.29$\\
    &$\theta_0^\prime=1.22$,\;$\theta_1^\prime=0.83$&$0.15$&$0.01$&$0.17$\\
\hline
    \end{tabular}
    \caption{$CHSH$ and $CHSH_E$ values for two sets of measurement parameters under all non-equivalent party assignments. Observables in zone 2 exhibit e-contextuality without c-contextuality; those in zone 3 are non-classical on both measures.}
    \label{tab:allfour}
\end{table}

\section{Conclusion}\label{sect:conclusion}
The question of allowed parameter values for state preparation and measurement has been previously raised in two contexts. On one hand, an axiom of quantum theory stipulates that these parameters vary continuously~\cite{hardy,grinbjps,Masanes,BruknerDakic}. On the other hand, continuous number fields, e.g. real numbers, may not be an adequate representation of physical reality~\cite{soler,Nielsen97,GisinReal}. We add a third aspect.

It is not surprising that the two fundamental criteria of non-classicality in the bipartite Bell scenario, viz. correlative and entropic CHSH inequalities, detect non-classical probability distributions differently. However, under a restriction on the continuous range of parameter values, there exists an anomalous discrepancy corresponding to e-contextuality \textit{without} c-contextuality. This result illustrates that specifying a value range, rather than admitting all theoretically allowed parameters, helps to precisely determine the type of non-classical resource.
Mathematically, the anomaly stems from the fact that entropic measures do not distinguish between perfectly correlated and anti-correlated observables. Conceptually, it connects the anomaly of entropic contextuality, via permutations, with the assignment of inputs and outputs to different parties in a device-independent approach.

Explored here through numerical simulation, the link between non-equivalent party assignments and imposing a restriction on parameter values should be amenable to experimental demonstration. For further work, it seems interesting to probe the zone of anomalous e-contextuality without c-contextuality empirically. To obtain a resource that remains non-classical under all non-equivalent party assignments, it would be equally interesting to implement the measurement settings that preserve non-classicality under all permutations.

%\bibliography{inf2}

\end{document}